\documentclass[12pt]{article}
\usepackage{amsmath,amsthm, amssymb}
\usepackage{graphicx}
\usepackage{leqno}
\usepackage{hyperref}
\numberwithin{equation}{section}
\begin{document}
\title{\vskip-40pt Is the Relativistic Structure of Spacetime Ontic or Epistemic?}
\author{Edward J. Gillis\footnote{email: gillise@provide.net}}

\maketitle

\begin{abstract}

\noindent 
As repeatedly emphasized by Einstein our knowledge of the structure of space and time is based entirely on inferences from observations of physical objects and processes. At the most fundamental level these objects and processes are described by quantum theory. However, the ontological status of the theoretical entities employed by quantum theory is a matter of considerable debate. Furthermore, the nonlocal correlations identified by Bell cannot readily be explained within the standard relativistic framework. This suggests that we should re-examine our understanding of spacetime.

\end{abstract}

\section{Can the Lorentzian metric structure of spacetime be considered complete? }
\label{sec:1}

\begin{quote}
	``All practical geometry is based upon a principle which is accessible to experience..." (A. Einstein)  \cite{Einstein_Geom}
\end{quote}

The usefulness of the standard relativistic description of spacetime is based on the premise that decipherable information about events in one region cannot be sent to other regions at a speed greater than that of light. The question that will be addressed here is whether the standard description takes into account \textit{all} of the fundamental features of spacetime.   In other words, are there hidden ontological structures that could help account for currently unexplained phenomena?\footnote{This inquiry will not deal with the very serious issues about the nature of spacetime that arise in connection with quantum gravity, cosmology, or extreme gravitational fields.}

Relativity was formulated by Einstein in a strictly classical context\cite{Einstein_1} roughly two decades before the development of quantum theory. The description of space and time in the new framework outlined by Einstein and formalized by Minkowski\cite{Minkowski} elegantly captured the intuitive notion that all causal influences propagate through space in a continuous manner. The incorporation of gravity into this scheme\cite{Einstein_GR} seemed to completely banish the spectre of action at a distance.

This intuitive notion about causal propagation has been labeled `local causality'. It has been given a fairly precise formulation by Bell\cite{Bell_LNC}. From it he derived an expression that describes the outcome probabilities of joint, spacelike-separated measurements, each with two possible outcomes. It relates these probabilities to instrument settings and other relevant physical quantities in the past light cone of those measurements. In a slightly modified notation this condition can be represented as: 
\begin{equation} \label{1x1} 
   Pr(A,B|a,b,c,\lambda) \; = \;   Pr(A|a,c,\lambda) \,   Pr(B|b,c,\lambda). 
\end{equation} 
In this expression $A$ and $B$ represent measurement outcomes; $a$ and $b$ indicate the physical quantities being measured (polarization in Bell's example); $c$ describes other features of the experimental arrangement; and $\lambda$ refers to any other relevant physical variables. The notation, $Pr$, refers to the probability of outcomes, $A$ and $B$, given the specification of $a,b,c,\lambda$. The factorization of the left side expression into the two independent terms on the right implies that the outcome, $A$, does not depend on either the setting, $b$, or the outcome, $B$, and that $B$ does not depend on either $a$ or $A$.

Bell further pointed out that quantum theory implies that the probabilities of spacelike-separated measurement outcomes \textit{do not} always obey \ref{1x1} (as demonstrated in his earlier work\cite{Bell_EPR}), thus making it clear that quantum theory is not locally causal. If one accepts Bell's methodological principle that systematic correlations require explanations the reasonable inference is that there are physical effects that are transmitted across spacelike intervals.

Einstein, along with Podolsky and Rosen, had already pointed out the conflict between local causality and quantum theory in 1935\cite{EPR}. They showed that a measurement on one of a pair of entangled particles that had separated to some distance would immediately reveal the state of the distant partner, even though this state was supposed to be undetermined prior to the measurement. They noted that one could measure either of two non-commuting observables, and argued that the supposedly uncertain quantities had to be previously determined.\footnote{The possibility of measuring either of two non-commuting observables is \textit{sufficient} to show that quantum theory requires the quantities to be undetermined prior to measurement, but there is debate on its exact role in the logical structure of the EPR argument. It is known that Einstein was unhappy with the way the argument was presented (presumably by Podolsky) in the published version. However, it is not necessary to go into all of the nuances of Einstein's thinking on this issue.} They concluded that quantum theory must be incomplete.

In his reply to the EPR paper Bohr simply talked around the issue that it had raised\cite{Bohr_Reply}. Since the application and further development of quantum theory did not exhibit any explicit contradictions with relativity the vast majority of the physics community ignored the key point that the EPR paper had made and adopted the attitude that Bohr had won the debate.

The reason that the nonlocal correlations do not conflict with the relativistic description of spacetime is, of course, that measurement outcomes obey the Born probability rule\cite{Born_prob}, which is enforced in quantum field theory by the requirement that spacelike-separated measurements commute. This was recognized by physicists who were forced to confront issues involving relativity. However, they regarded this requirement as a straightforward adaptation of the idea that the speed of light is an absolute limit. In his critique of this way of thinking Bell\cite{Bell_LNC} quotes a passage from Gell-Mann, Goldberger, and Thirring\cite{GMGbTh}: 
 \newline 
\begin{quotation}
	\noindent 
	``The quantum mechanical formulation of the demand that waves do not propagate faster than light is, as is well known, the condition that the measurement of two observable quantities should not interfere if the points of measurement are space-like to each other...the commutators of two Heisenberg operators...shall vanish if the operators are taken at space-like points." 
\end{quotation} 
This requirement is known as `local commutativity'. 

In fact, the condition of local commutativity is quite different from the claim that wave propagation is limited by the speed of light. This can be seen in several ways. First, it is necessary to assume the invariance of the speed of light in order to define `local commutativity' since the notion of spacelike separation depends on it. 
This invariance alone made it possible to construct classical theories such as general relativity and Maxwell's electromagnetism in which the speed of light is an upper limit. The need to add the postulate of local commutativity to maintain the Lorentz invariance of quantum field theory strongly suggests that this new theory introduces novel effects that must be controlled. Second, it is simply not true that propagation is limited by the speed of light in quantum field theory (as pointed out by Peskin and Schroeder\cite{Peskin_Schroeder}):
\begin{quote}  
``Thus, the amplitude is small but nonzero outside the light cone, and causality is still violated." (Peskin and Schroeder, p.14) 
\end{quote} 
Most importantly, the nonlocal correlations identified by Bell cannot be explained without nonlocal physical effects. The reason for assuming local commutativity is to prevent these superluminal effects from transmitting decipherable information.

In classical theory the limiting speed of light prevents physical processes from propagating outside the light cone, and beautifully captures our pre-theoretic intuition that causal influences travel through space in a continuous manner. By tying this deep intuition to the relativistic structure of spacetime it served to cement the conviction that the relativistic account is an accurate and \textit{complete} description of fundamental spacetime ontology. Because the postulate of local commutativity prevents the transmission of decipherable information at superluminal speeds it guarantees the Lorentz invariance of quantum field theory and makes possible the continued use of the relativistic description of spacetime as a framework for that theory. The solid conviction that relativity accurately characterized spacetime ontology led many people to conflate the two very different concepts.

Because the ontological status of the quantum wave functions and operators remained obscure it was easy for most working physicists to ignore the foundational questions, in particular those regarding the relationship of quantum processes to spacetime structure. Furthermore, the perfect correlations (or anticorrelations) pointed out in the EPR paper do not themselves conflict with local causality, even though they are inconsistent with the claim that quantum theory is both locally causal and complete. Although there were a few people in addition to Einstein who were bothered by some of the nonlocal aspects of quantum theory it was only when Bell showed an explicit contradiction between some quantum correlations and local causality in \cite{Bell_EPR} that the problem was brought into sharp focus. The experimental confirmation of the existence of these problematic correlations by Aspect\cite{Aspect_1,Aspect_2} helped to induce some to more seriously question the idea that the relativistic description of spacetime is complete.

A number of researchers in addition to Bell have made serious attempts to sort out the issues involved. Maudlin has examined the problem at length in several works\cite{Maudlin_1,Maudlin_2,Maudlin_3}. In \cite{Maudlin_2} he offers ``the proper definition of a Relativistic theory" as ``a theory that postulates only the Relativistic metric as space-time structure." He also notes that ``the light-cone structure is determined by the Relativistic metric". Elsewhere he argues that for any theory to be regarded as `fundamentally relativistic' it must not rely essentially on any additional spacetime structure. While this last remark can be legitimately debated, this viewpoint will be adopted here since it serves to frame the question of what features one might add to the metric in order to explain the nonlocal correlations described by Bell. 

An alternate perspective has been outlined by D\"{u}rr, and coauthors in ``Can Bohmian mechanics be made relativistic?"\cite{Bohm_Rel} They acknowledge that ``privileged space-time foliations" are an essential ingredient in Bohmian theories, but they argue that these could be extracted from the wave function ``so that a theory
formulated in this way should be regarded as fundamentally Lorentz invariant." Although they diverge from Maudlin in their opinion of what should be regarded as fundamentally relativistic, the critical point for the issue under consideration here is the recognition that a foliation of spacetime is needed to complement the metric in Bohmian theory.\footnote{They also briefly describe how a generalization of Bohmian theory could be based on a time-like vector field on spacetime. Such a field can be defined in relativistic spacetime without postulating additional structure. If the field is integrable, a foliation can be derived from it.}

An interesting and somewhat different approach has been described by Gisin and his colleagues. In \cite{Bancal} it was shown that any proposal to account for the nonlocal correlations by superluminal influences that propagate at \textit{finite} speeds (no matter how large) would lead to signalling. They suggest ``that quantum correlations somehow arise from outside spacetime, in the
sense that no story in space and time can describe how they occur." Gisin expands on this suggestion slightly with the idea that ``one random event can manifest itself at several locations"\cite{Gisin_no_finite}. While this proposal does not, strictly speaking, add structure to relativistic spacetime, it certainly expands the framework of theoretical physics beyond the standard relativistic perspective.

Another strategy for explaining the nonlocal effects is to modify the Schr\"{o}dinger equation by adding nonlinear, stochastic terms. The additional terms are designed to bring about a nonlocal collapse of the wave function in accord with the Born probability rule. The nonlinearity is necessary to produce an actual collapse (in contrast with Bohmian theory which attempts to explain the \textit{appearance} of collapse). The stochastic nature of the equation is dictated by the fact that any deterministic nonlinear extension of the Schr\"{o}dinger equation would violate the prohibition against superluminal signaling, as demonstrated by Gisin\cite{Gisin_c}. 

This general approach has been implemented in a number of proposals in both relativistic and nonrelativistic settings. In a recent work  Tumulka\cite{Tumulka_Rel_Interact} presents what he describes as ``a relativistic model of spontaneous wave function collapse for N distinguishable particles with interaction".\footnote{Tumulka's paper builds on previous work that he and several others have done on developing relativistic collapse models. See the sources cited in Tumulka's paper.} The model does not involve any preferred foliation of spacetime, and is, therefore, fundamentally relativistic in the sense advocated by Maudlin. Although there are some limitations clearly described by Tumulka, the demonstration of the compatibility of wave function collapse with such a spacetime is a substantial achievement.

The efforts of Tumulka and others in this pursuit can be described as an attempt to reconcile quantum theory \textit{with} relativity. Essentially, from this perspective the fundamentally relativistic framework is taken as given, and the task is to find a way to fit quantum theory into it. This is a reasonable attitude since relativity provides a fairly simple and elegant description of spacetime, and since it imposes strong constraints on acceptable theories; and, of course, the foundations of quantum theory are rightly regarded as being more obscure. 

However, there is another perfectly reasonable approach to the problem of establishing the compatibility of these ``two fundamental pillars of contemporary theory"\cite{Bell_fund}. One could try to reconcile quantum theory \textit{and} relativity. In other words, these two pillars should be regarded as being equally ``fundamental", and as being equally open to reasonable modifications. This is the tack that has been taken by this author in a several works\cite{Gillis_1,Gillis_2,Gillis_3}.

To pursue this line of attack one begins by asking what is the fundamental feature that these two theoretical frameworks have in common. It is the prohibition of superluminal signaling.\footnote{To be precise, to derive this prohibition from the fundamentally relativistic structure of spacetime, one must add the assumption that signals cannot be sent into the past in any reference frame. This assumption is implicit in most theories.} Bell acknowledged this point, but he was very pessimistic that it could be used to help resolve the apparent conflicts between relativity and quantum theory:  
\begin{quote}
	``Do we then have to fall back on `no signalling faster than light' as the expression of the fundamental causal structure of contemporary theoretical physics? That is hard for me to accept. For one thing we have lost the idea that correlations can be explained, or at least this idea awaits reformulation. More importantly the 'no signalling...' notion rests on concepts that are desperately vague or vaguely applicable."
\end{quote}
He appeared to regard local commutativity (with its implied lack of determinism) as a possible key to understanding the basis for the no signalling principle. He went on to add that the challenge is to ``couple it with sharp internal concepts, rather than vague external ones."

Bell provides a fairly precise characterization of `local causality', but he gives only a very general gloss of what a `fundamental causal structure' involves. Apparently it must insure that ``correlations can be explained". In the classical theories of electromagnetism and general relativity the postulate of the invariance of the speed of light, along with the assumptions of the equivalence of reference frames and the principle that causes precede effects, insures that all physical influences propagate within the light cone, and, thus, all systematic correlations can be readily explained. In these theories the fundamental causal structure coincides with the notion of local causality.

To explain the nonlocal correlations that are implied by quantum theory we need to assume that there are causal influences that act between entangled systems across spacelike intervals. This entails that the relativistic metric no longer completely determines the fundamental causal structure of the theory. Note that this is true even in explanations of wave function collapse like that of Tumulka that do not add any structure to the relativistic metric. This could be regarded as reducing the motivation for trying to remain strictly within the relativistic framework. The justification for adding something like a foliation to spacetime will depend on whether such an addition leads to a simpler or more elegant account.

Any such account must, of course, prevent superluminal signaling and be Lorentz invariant at the level of observations. The proposal presented in \cite{Gillis_3}, besides meeting these minimal requirements, is able explain wave function collapse \textit{without assuming any new physical constants}, and it is also \textit{consistent with strict adherence to conservation laws in individual cases}. These points will be expanded upon below; the full account is given in the work cited.

As indicated above the starting point for the construction of the account in \cite{Gillis_3} was the identification of the fundamental feature shared by quantum theory and relativity. This is the prohibition of the transmission of reproducible, decipherable information across spacelike intervals. To implement this sort of approach it is necessary to develop an explanation for wave function collapse in terms of fundamental physical processes that has this prohibition as a logical consequence. This does not require that such an explanation must completely define `information' in strictly fundamental terms. It is sufficient to identify the fundamental processes that are critical to the physical instantiation and transmission of information

The fundamental processes that are absolutely essential to sending signals or transmitting reproducible information are interactions that establish correlations between the states of the interacting systems. These correlating interactions provide the key to developing an account of wave function collapse that has all of the features described in the previous few paragraphs.

The basic idea is quite straightforward. One assumes that every correlating interaction induces a small transfer of amplitude between the interacting and noninteracting components of the systems involved in the interaction. The direction of the transfer is determined by a single global stochastic process that is tied to a preferred foliation of spacetime. With enough such transfers the wave function collapses to one of the entangled systems that are generated by the correlating interactions. It is worth noting that all stochastic collapse proposals rely on such large scale entanglement to be effective. This point is sometimes obscured by the emphasis on the collapse of an individual elementary target system. The large scale entanglement of the systems plays a critical role in insuring strict consistency with conservation laws.

The simplest way to implement a collapse proposal that involves a foliation of spacetime is to assume a preferred inertial frame. Since the nonlocal quantum effects appear in nonrelativistic versions of the theory it is possible to illustrate the essential features of the proposal while avoiding most of the complications of quantum field theory. Interactions are modeled with two-particle interaction potentials, and the use of configuration space as the mathematical setting for the collapse equation simplifies the description of the relevant entanglement relations and the collapse process. It also helps to make clear that the collapse involves an entire entangled branch of the wave function.

The stochastic operator is based on two-particle interaction potentials. These are assumed to be conservative and distance-dependent. The distance-dependent nature of the potentials is responsible for the localization of the collapses, and it also eliminates the need for a new physical constant to set the range of the effects. The magnitude of the amplitude shift associated with each interaction is determined by the ratio of interaction potential energy to total relativistic energy. In nonrelativistic situations these magnitudes are generally about $10^{-4}$ or smaller. The small size of the individual shifts makes it possible for the stochastic action to occur in parallel with the ordinary Schr\"{o}dinger evolution determined by the Hamiltonian.

The timing parameter is determined by the rate at which correlations are generated between the interacting systems. When systems interact they tend to settle into a stationary state or separate to a distance at which the interaction effectively ceases in a very short time. For the strongest nonrelativistic interactions this time is on the order of $10^{-17}$ seconds. The timing parameters are defined in terms of the rate at which potential energy is converted to kinetic energy (or vice versa), divided by the total nonrelativistic energy as measured in the center-of-mass frame of the interacting systems. The integrated value of the timing parameters over the period during which the correlation is established is approximately $1$ for those interactions that are most effective in inducing collapse.

By basing the collapse process on correlating interactions and recognizing the entanglement that invariably results from interactions it is possible to insure consistency with major conservation laws. This consistency depends on the fact that even elementary systems that are prepared in supposedly ``factorizable" states are entangled to some extent with the preparation apparatus. When the elementary systems are ``measured" the collapse that is induced also affects the state of the preparation apparatus. Although the differences in the states of the apparatus associated with various measurement outcomes are minuscule, they are sufficient to compensate for the perceived violation of conservation laws associated with the measured system.\footnote{Consider a photon that passes through a beam-splitter and is then detected in the reflected branch. The apparent violation of momentum conservation in the reflected branch is offset by a tiny change in the state of the beam-splitter.}

Specifically, the proposal in \cite{Gillis_3} is able to insure complete consistency with the conservation of momentum and orbital angular momentum. Consistency with energy conservation is maintained within the accuracy that can be achieved, given the types of particles recognized by nonrelativistic theory. (Since small contributions from photons, relativistic corrections to kinetic energy, and even antiparticles occur in the kinds of interactions that induce collapse one cannot expect an exact energy accounting.)  

The equation based on the idea that wave function collapse is induced by correlating interactions also implies the Born probability rule. This prevents superluminal signaling. It also insures that the order of spacelike-separated events (including measurements) is completely undetectable. Thus, there is no way to determine the preferred rest frame. On an observational level all inertial frames are equivalent, and so Lorentz invariance is maintained in descriptions of spacetime.

Clearly, the assumption of a preferred frame (or more general foliation) violates fundamental relativity in the sense advocated by Maudlin. The question is whether this feature outweighs the advantages of an account such as that presented in \cite{Gillis_3}. These advantages include the elimination of the need to introduce new, ad hoc physical constants and strict consistency with conservation laws. The only new feature that it assumes (in addition to the preferred foliation) is a single, global stochastic process. By providing a more complete ordering of events in spacetime the preferred foliation also insures that all relevant entanglement relations are well defined when collapse occurs.

To address the question posed by the title of this paper let us first consider the ontological status of quantum states. This status was the subject of a work by Pusey, Barrett, and Rudolph\cite{PBR} in which they showed:
\begin{quote}
``that any model in which a quantum state represents mere information about an underlying physical state of the system, and in which systems that are prepared independently have independent physical states, must make predictions that contradict those of quantum theory."
\end{quote}
In other words, the quantum state is not a strictly epistemic entity. It is, at least in part, ontic. 

Nevertheless, quantum theory, at the level of observation, is probabilistic. The wave functions that represent quantum states in nonrelativistic theory live in configuration space - not in physical space. Which quantum state applies in a given situation depends on whether the state gets updated when ill-defined measurements take place. These considerations strongly suggest that these states have both ontic and epistemic features.

What about the ontological status of relativistic spacetime? As clearly argued by Maudlin the relativistic metric determines the light cone structure, and this plays a critical role in both classical and quantum theories. Hence, any modifications to our  understanding of spacetime will be additions to the relativistic structure. So this structure must be, at least in part, ontic. How strong then is the case that it should also be regarded, to some extent, as epistemic?

Various approaches to explaining how nonlocal quantum effects are related to relativistic spacetime have been briefly reviewed above. Bohmian theory requires a foliation of spacetime. The authors of \cite{Bohm_Rel} propose that this foliation can be derived from the wave function defined on some spacelike surface. They argue that this should be considered as ``fundamentally relativistic" because the foliation is derived rather than simply postulated. 

The results described in \cite{Bancal,Gisin_no_finite} set a strong constraint on any proposed explanation. It is shown that the nonlocal effects must appear to be transmitted instantaneously in every reference frame. This is essentially equivalent to the requirement that Lorentz invariance must be maintained \textit{at the level of observation}. The authors suggest ``that quantum correlations somehow arise from outside spacetime". It would be necessary to offer a specific proposal about how this might be done in order to judge whether it would count as fundamentally relativistic.

The work of Tumulka and others cited in \cite{Tumulka_Rel_Interact} shows that it is possible to embed an account of nonlocal collapse in a strictly relativistic spacetime, but these proposals come at a cost. It is necessary to postulate new physical constants in a rather ad hoc manner, and there are some violations of conservation laws.

In \cite{Gillis_3} it is shown how these costs can be avoided if one is willing to add a foliation to spacetime. So let us assess the trade-offs involved in modifying physical theory in this manner.

One of the primary reasons that many physicists believe that the relativistic metric should be considered as completely describing spacetime ontology has already been addressed. The intuitive idea that all physical processes propagate through space in a continuous manner is given a deep explanation in terms of spacetime structure. However, these intuitions are based on our macroscopic experience, in particular, the fact that we have no control over effects that are not continuously propagated. In the light of Bell's demonstration that there are real nonlocal effects, this motivation can no longer be seen as a compelling reason to reject additional spacetime structure.

Other possible motivations for trying to avoid adding structure to spacetime were mentioned briefly above: (1) the simplicity and elegance of the relativistic description, (2) the strong constraints that this description places on potential theories, and (3) the fact that the current obscurity of the foundations of quantum theory should make one reluctant to fundamentally alter our view of spacetime.

The third of these considerations, regarding the obscurity of standard quantum theory, is exactly what the proposals that would add spacetime structure are intended to deal with. The first and second are both tied to the assumption that the ontology of spacetime is completely described by the Lorentzian metric which, in turn, defines the light cone structure and imposes the requirement that acceptable theories should be Lorentz invariant. For this reason let us focus on the role of Lorentz invariance in contemporary theory.

It is enlightening to trace Bell's thinking on this matter. Bell used the term ``fundamentally Lorentz invariant" as essentially equivalent to Maudlin's phrase, ``fundamentally Relativistic". He acknowledged that additional spacetime structure such as a preferred foliation could be used to explain the nonlocal quantum correlations, but appeared to indicate a preference to avoid any such additions. In discussing the de Broglie-Bohm theory in \cite{Bell_Qmfc} he entertained the possibility that Lorentz invariance holds only at the level of observations:
\begin{quote}
``It may well be that a relativistic version of the theory, while Lorentz invariant and local at the observational level, may be necessarily non-local and with a preferred frame (or aether) at the fundamental level."
\end{quote} 
However, even though he had praised the pilot wave theory for helping to clarify some of the key foundational issues, he later expressed considerable skepticism that it could provide a final resolution to the issues\cite{Bell_Lor_Inv}: 
\begin{quote}
	``For me, this is an incredible position to take - I think that it is quite logically consistent, but when one sees the power of the hypothesis of Lorentz invariance in modern physics, I think you just can't believe in it." 
\end{quote}

This skepticism might have been inspired, in part, by his analysis of the stochastic collapse proposal of Ghirardi, Rimini, and Weber\cite{GRW} which he suggested opens the door to a fundamentally relativistic account\cite{Bell_Atqj}:
\begin{quote} 
``that the model is as Lorentz invariant as it could be in the nonrelativistic version. It takes away the ground of my fear that any exact formulation of quantum mechanics must conflict with fundamental Lorentz invariance."	
\end{quote}
He also made brief reference to the possibility of maintaining fundamental Lorentz invariance with a stochastic collapse model in \cite{Bell_LNC}:
 \begin{quote}
 	``Perhaps there is a already a hint of this in `quantum mechanics with spontaneous wavefunction collapse."
 \end{quote}

There is no denying ``the power of the hypothesis of Lorentz invariance in modern physics". However, the question that can be raised is at what level is it necessary to invoke this hypothesis. It is used to great effect in quantum field theory. But, Bell, himself, (who had derived key features of quantum field theory from it\cite{Bell_Time_Rev}) did not appear to regard the theory in its current form as an ``exact formulation". Apparently, in his view it describes things only at the level of observation. Thus, the usefulness of the hypothesis does not seem to require that it completely characterizes the fundamental ontology of spacetime. It would seem that requiring Lorentz invariance at the observational level could both constrain the construction of conventional quantum theories and guide the development of more fundamental accounts.

The work of Tumulka and others has shown the possibility of constructing a fundamentally relativistic account that explains the nonlocal correlations. However, alternative accounts that invoke foliations such as the pilot wave theory and the stochastic collapse proposal outlined in \cite{Gillis_3} have shown that the nonlocal effects can be explained without violating conservation laws or postulating new physical constants. In these alternate accounts the preferred foliation is, in principle, undetectable. Hence, they maintain Lorentz invariance at the level of observations.

Since quantum theory as it is currently formulated is probabilistic, (and since this lack of determinism is mirrored in stochastic collapse proposals) one can reasonably ask why should we expect that all fundamental structures of spacetime should be revealed by observation. The probabilistic nature of the theory could shield certain ontological features of spacetime from our view, just as it shields some aspects of quantum wave functions. 

Given these considerations, it is quite reasonable to consider the possibility of such hidden structures, and to weigh the trade-offs between accounts that include them and those that adhere to the pre-quantum view of relativity theory.

\end{document}